\documentclass[12pt]{article}
\usepackage[pdftex]{graphicx}
\usepackage{amssymb}
\usepackage{amsmath}

\newcommand{\rf}[1]{(\ref{#1})}
\newcommand{\beq}{\begin{equation}}
\newcommand{\eeq}{\end{equation}}
\newcommand{\bea}{\begin{eqnarray}}
\newcommand{\eea}{\end{eqnarray}}

\newcommand{\e}{\mbox{e}}

\newcommand{\lam}{\lambda}
\newcommand{\Lam}{\Lambda}

\newcommand{\m}{\mu}

\newcommand{\ep}{\varepsilon}

\newcommand{\sg}{\sigma}

\newcommand{\oh}{\frac{1}{2}}

\newcommand{\tr}{\mathrm{tr}\,}
\newcommand{\ra}{\rangle}
\newcommand{\la}{\langle}
\newcommand{\prt}{\partial}

\newcommand{\cT}{{\cal T}}

\newcommand{\tZ}{{\tilde{Z}}}

\begin{document}

\begin{center}
\vspace{24pt}
{ \large \bf New multicritical matrix models and multicritical 2d CDT}

\vspace{30pt}

{\sl J. Ambj\o rn}$\,^{a}$,
{\sl L. Glaser}$\,^{a}$,
{\sl A. G\"{o}rlich}$\,^{a,b}$
and {\sl Y. Sato}$\,^{a,c}$

\vspace{48pt}
{\footnotesize

$^a$~The Niels Bohr Institute, Copenhagen University\\
Blegdamsvej 17, DK-2100 Copenhagen \O , Denmark.\\
email: ambjorn@nbi.dk, glaser@nbi.dk, goerlich@nbi.dk, ysato@nbi.dk\\

\vspace{10pt}

$^b$~Institute of Physics, Jagiellonian University,\\
Reymonta 4, PL 30-059 Krakow, Poland.\\
email: atg@th.if.uj.edu.pl\\

\vspace{10pt}

$^c$~Department of Physics, Nagoya University, \\
Nagoya 464-8602, Japan.\\
email: ysato@th.phys.nagoya-u.ac.jp

}
\vspace{96pt}
\end{center}


\begin{center}
{\bf Abstract}
\end{center}

We define multicritical CDT models of 2d quantum gravity
and show that they are a special case of multicritical 
generalized CDT models obtained from the new scaling 
limit, the so-called ``classical'' scaling limit, of matrix models.
The multicritical behavior agrees with the multicritical behavior 
of the so-called branched polymers.

\noindent 
\vspace{12pt}
\noindent

\vspace{24pt}

\newpage

\section{Introduction}\label{intro}
 
Multicritical matrix models have served as an important tool 
in the study of 2d quantum gravity coupled to matter. They 
were introduced by Kazakov for this purpose \cite{kazakov}
and the identification of the first multicritical matrix
model limit as corresponding to a $(p,q)=(2,5)$ minimal 
rational conformal field theory coupled to 2d Euclidean 
quantum gravity was due to Staudacher \cite{staudacher}.

The concept of multicritical so-called branched polymers (BP)
was introduced in an attempt to understand the physics 
underlying the multicritical matrix models \cite{adj-bp}. 
Ordinary branched polymers were encountered in the theory 
of random surfaces which aimed  at providing a non-perturbative
definition of the Polyakov path integral, either on 
a hyper-cubic lattice \cite{djf} or using the 
formalism of dynamical triangulations (DT) \cite{DT}.
For these non-perturbatively defined random surface
theories the result was that in physical dimensions 
($D \geq 2$) the random surfaces degenerate into branched
polymers \cite{djf,ad}. The same phenomenon was 
observed in attempts to study higher dimensional Euclidean 
quantum gravity using DT \cite{higherD}. Branched polymers 
seem to be very generic structures which are entropically 
favored and in statistically inspired models of random geometry
one has to make special efforts to avoid them. 

The formalism of  causal dynamical triangulations (CDT) represents
an attempt to eradicate the dominance of BP. It has
been partly successful as a model of higher dimensional 
quantum gravity \cite{CDT-highD}. The two-dimensional 
CDT model \cite{al} is special in the sense that it is exactly solvable
and in the sense that it can be understood as a specific 
limit of ordinary Euclidean quantum gravity where 
baby universes have been integrated out \cite{ackl}.
It is somewhat amusing that while one of the 
purposes of the CDT model was to avoid BP, the two-dimensional
model is entirely described by BP, as first noticed in \cite{djw}.
We can use the BP description of two-dimensional CDT in a 
constructive way to formulate multicritical CDT models.
In the following we will do that, and we will show that these 
multicritical CDT models are special cases of more general 
multicritical matrix models. The scaling limit taken 
for these matrix models, the so-called classical scaling limit,
is different from the conventional scaling limit referred to 
above. This new scaling limit was first introduced in a matrix model
with the purpose of obtaining CDT from a matrix model \cite{newmatrix}. We 
present here the generalization to multicritical behavior.

\section{Multicritical branched polymers and multicritical CDT}   

\subsection{Multicritical branched polymers}
      
One can define a statistical ensemble of BP by the partition function
\beq\label{1.1}
Z(\m) = \sum_{BP} \prod_{i} v(i) \prod_l \e^{-\m}.
\eeq
The sum is over graphs which are rooted connected planar trees.
The root is a distinguished vertex. For simplicity we assume 
there is only one link incident on the root. This assumption
has no consequences for the critical behavior of the ensemble
of BP. To each vertex $i$ is assigned a weight $v_i$ which we usually 
take to depend only on the order of the vertex. To each link
$l$ is assigned a fugacity $\e^{-\m}$. 

\begin{figure}[t]
\centerline{\scalebox{1.0}{\includegraphics{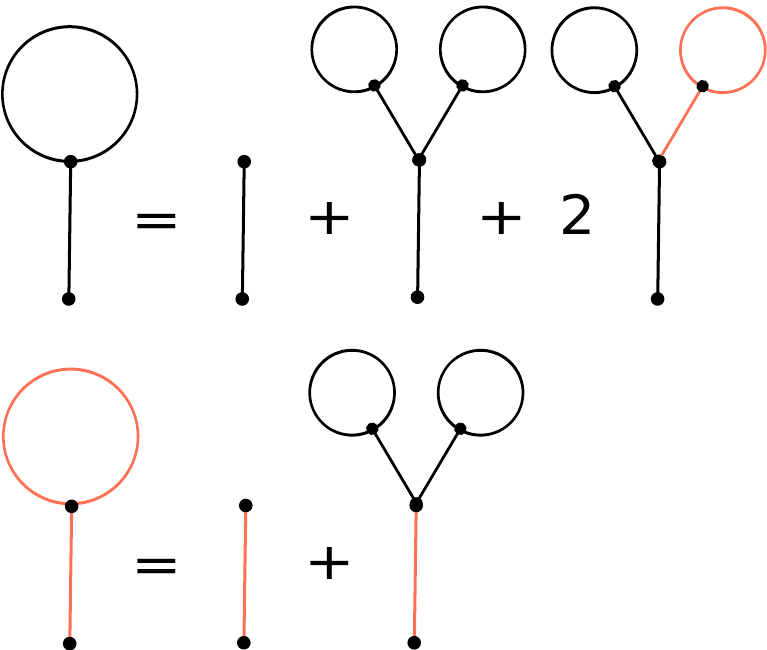}}}
\caption{The graphical representation of eqs.\ \rf{1.6}.
The black colored graphs represent the class of 
BPs where the link touching the root contains no dimer.
The red-colored graphs represent the 
class of BPs where a dimer touches the root. It is assumed
that only one link is incident on the root vertex. This 
restriction can easily be lifted, but is imposed as it leads 
to slightly simpler algebra.}
\label{fig1}
\end{figure}

The graphical selfconsistent equation which determines 
$Z(\m)$ is shown as the lower part of Fig.\ \ref{fig1}, with the 
colors ignored, in the special case where $v_1=v_3=1$, all other 
$v_i$ being zero. 

It leads to the 
following equation
\beq\label{1.2}
\e^{\m} = \frac{1+v_2 Z +v_3 Z^2 \cdots }{Z} := \frac{f(Z)}{Z} :=F(Z).
\eeq 
We assume by simple rescaling that $v_1=1$.
The generic BP is obtained if the weights $v_i$ are non-negative\footnote{One
can obtain non-generic behavior for non-negative weights $v_i$ if 
infinite many  $v_i > 0$ and  $F(Z)$, $Z >0$, is a monotonic decreasing 
function in the range  where it is defined \cite{adf3}.}.
In this case $F(Z)$ has an minimum $Z_0$ where $F'(Z_0)=0$ and 
$F''(Z_0) >0$. Thus we have the non-analytic behavior
\beq\label{1.3}
Z(\m)-Z(\m_0) \sim (\m-\m_0)^{1/2},~~~e^{\m_0} = F(Z_0),
\eeq
in the neighborhood of the critical point $\m_0$.

However, if we give up the requirement that the weights $v_i$ of branching
should be positive it is clear that there exists special choices of 
the $v_i$ such that not only $F'(Z_0)=0$, but also $F^{(k)}(Z_0)=0$,
$k=1,\ldots m-1$. In this case we obtain
\beq\label{1.4}
Z(\m)-Z(\m_0) \sim c\,(\m-\m_0)^{1/m},~~~e^{\m_0} = F(Z_0).
\eeq
For $m>2$ we say that the ensemble of BP is multicritical 
\cite{adj-bp,book,BB}. 
The fractal structure of the BP depends on $m$. 
The Hausdorff dimension of a type $m$ multicritical BP is 
\beq\label{1.5}
d_h (m) = \frac{m}{m-1}.
\eeq
As $m$ increases the BP gets more and more one-dimensional 
at the critical point.

The multicritical behavior of the BP can be given a concrete
realization as hard dimer models with negative fugacities 
on BPs with positive weight, very 
much like the situation for the multicritical matrix models,
as pointed point in \cite{bonzom}.
We define a BP hard dimer model as
\beq\label{1.8}
Z(\m,\xi) = 
\sum_{BP}  \prod_{i} v(i) \prod_l \e^{-\m} \sum_{{\rm HD(BP)}} 
\xi^{|{\rm HD(BP)}|}.
\eeq
Here we will assume all $v_i$ non-negative and for each BP
we sum over all hard dimer configurations. The dimers 
live on links in the BP,  ``hard'' meaning 
that the dimers are not allow to touch. To each dimer we 
associate a fugacity $\xi$. For a given dimer configuration $HD$
the total weight will thus be $\xi^{|HD|}$, $|HD|$ denoting
the number of dimers in the configuration $HD$. 
To illustrate this let us consider the simplest BP model 
with positive weight, namely $v_1,v_3=1$, all other weights 
equal zero. Thus $F(Z)=(1+Z^2)/Z$ and the model 
indeed has  a critical point, $\m_0=\log (2),Z_0=1$. Now allow hard
dimers on the links, and let the fugacity of a dimer be $\xi$. 
As shown in Fig.\ \ref{fig1} we have now two rooted partition functions, one
starting with a link without a dimer ($Z(\m,\xi)$), the other starting
with a link with a dimer ($\tZ(\m,\xi)$). The corresponding equations are:
\beq\label{1.6}
\e^\m = \frac{1 +Z^2 +2Z \tZ}{Z}, ~~~\e^{\m}= \xi \; \frac{1+Z^2}{\tZ}.
\eeq
For positive fixed values of $\xi$ one has the standard critical 
behavior of $Z$ as a function of $\m$, i.e. a critical point
$\m_0(\xi)$ such that near that point \rf{1.3} is valid.
However, writing \rf{1.3} as 
\beq\label{1.6a}
\m-\m_{0}(\xi) = c_2(\xi) \Big(Z(\m)-Z(\m_0(\xi))\Big)^2+ c_3(\xi) 
\Big(Z(\m)-Z(\m_0(\xi))\Big)^3 +\cdots,
\eeq  
the coefficient $c_2$ will decrease as $\xi$ becomes negative,
and eventually one will reach a point  $\xi_c$, with a 
critical point $\m_c= \m_0(\xi_c)$ where $c_2(\xi_c)=0$.
At the  corresponding $Z_c= Z_0(\m_c,\xi_c)$ we have 
\beq\label{1.7}
\frac{\prt \m(Z,\xi_c)}{\prt Z}\Big|_{Z_c} =  
\frac{\prt^2 \m(Z,\xi_c)}{\prt Z^2}\Big|_{Z_c} = 0
\eeq
Clearly this point is a $m=3$ multicritical point:
\beq\label{1.8a}
\m-\m_c \sim c_3 (Z(\m,\xi_c)-Z_c)^3,~~~{\rm i.e.}~~~ 
Z(\m,\xi_c)-Z_c \sim c (\m-\m_c)^{1/3}. 
\eeq

 One can  calculate the critical exponent
related to dimers, i.e. the matter system. Following \cite{kazakov,staudacher}
the critical $\m_0(\xi)$ has the interpretation
as the free energy density of the matter system and the critical 
exponent $\sg$ (corresponding to the magnetization at the critical 
temperature in the case of a magnetic system) can be defined as 
\beq\label{1.9}
\frac{\prt \m_0(\xi)}{\prt \xi} \sim (\xi-\xi_c)^\sg
\eeq
where the term on the rhs denotes the leading non-analytic term
when $\xi$ approaches $\xi_c$.
It is straight forward to show that in addition to \rf{1.8a}
we have
\beq\label{1.10}
\xi-\xi_c\sim (Z(\m_0(\xi),\xi)-Z_c)^2,
\eeq
from which we deduce that $\sg =1/2$.

\begin{figure}[t]
\centerline{\scalebox{1.0}{\includegraphics{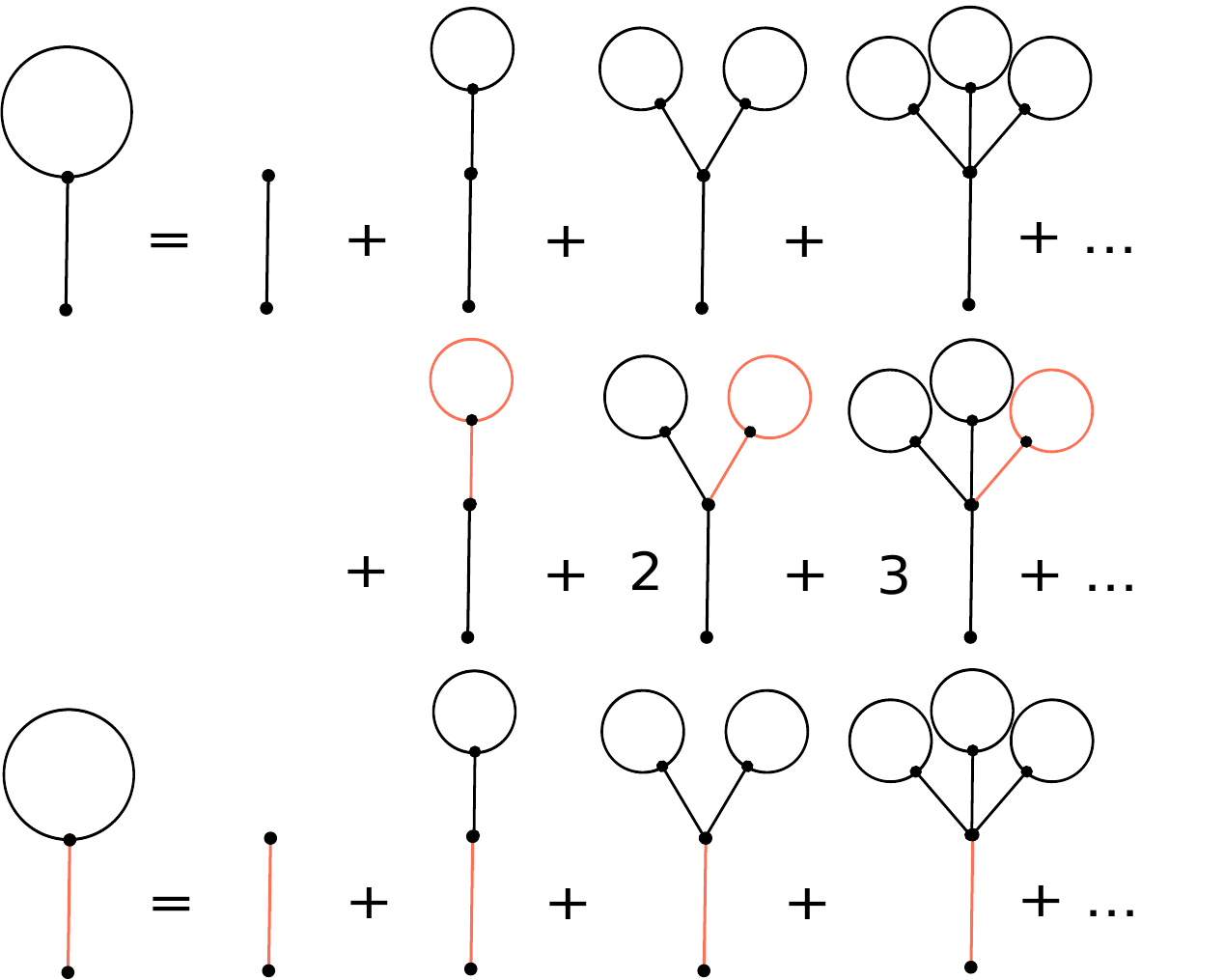}}}
\caption{The graphical representation of the equations \rf{1.10a}.}
\label{fig2a}
\end{figure}

The multicritical behavior reported above in terms of dimers 
is generic for BPs and independent of the particular model.
Another explicit BP model, relevant for the theory of causal 
dynamical triangulations, is one where branching of arbitrary high
order, with weight one, is allowed. The graphical equations are shown 
in Fig.\ \ref{fig2a} and the equations corresponding to \rf{1.6} are:
\beq\label{1.10a}
\e^\m = \frac{1}{Z} \left( \frac{1}{1-Z} +\frac{\tZ}{(1-Z)^2}\right),~~~~
\e^\m \tZ = \xi \frac{1}{1-Z}.
\eeq
Also this model can be solved explicitly and the critical behavior is 
as above.

The simplest dimer generalization to higher multicritical points
is obtained by considering several kinds of dimers with independent 
fugacities and allow them to touch the same vertex, but not 
to be put on the same link. One can then arrange for the
fugacities such that one obtains a $m=4$ multicritical BP.

We note the following: an Ising model coupled to a generic BP cannot be
critical \cite{durhuus}. Although the Hausdorff dimension
of the BP is two, part of the inherent structure of the 
BP is still too one-dimensional to allow for a critical 
behavior. Thus Ising spins cannot change the fractal 
properties of an ensemble of BP. However the above calculation shows
that coupling hard dimers to BPs can lead to a change of the 
critical properties of branched polymers. We will  comment on
the criticical properties of the dimers below.  

\subsection{Multicritical CDT}

\begin{figure}[t]
\centerline{\scalebox{1.2}{\includegraphics{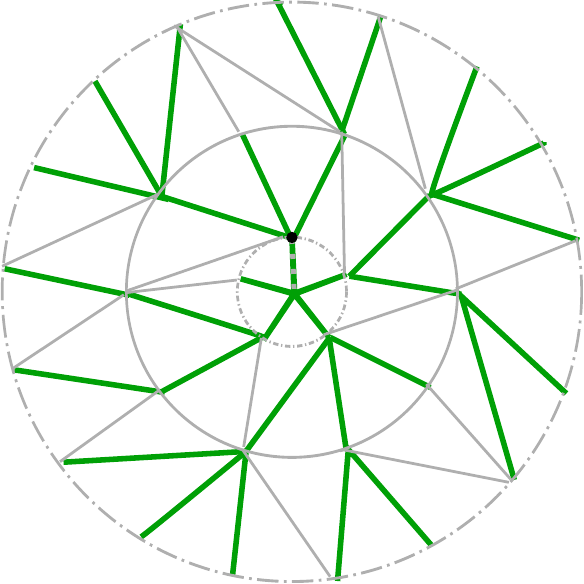}}}
\caption{The construction of a tree on a CDT configuration. The inner
circle is the entrance boundary. The interior to the 
inner circle is just an additional construction to obtain bijection
between the ensemble of BPs and the CDT ensemble of random
surfaces (see \cite{djw} for a discussion).}
\label{fig2}
\end{figure}

The ordinary  two-dimensional CDT theory is 
defined by the following partition function
\beq\label{1.11}
Z_{CDT}(t) = \sum_\cT \e^{-\m N(\cT) },
\eeq
where the summation is over the set of so-called causal 
dynamical triangulations $\cT$, usually defined with 
two boundaries, separated a link distance $t$. 
We might contract the boundaries to contain only a point 
if convenient. For 
detail we refer to \cite{al}. There is a bijection between 
causal triangulations and BPs, as first observed 
in \cite{djw}. Let us mention how this assignment 
is done. To each vertex in a CDT triangulation, except those
at the two boundaries we have a number of links pointing forward
in time, a number of links pointing backwards in time and two spatial 
links. The assignment is now that  all links pointing forward
from a vertex, except one link which one can take as the link to
the left, belong to the BP.  
For the entrance boundary, which has one marked vertex, a
special assignment has to be made. For details we refer to \cite{djw}
and for an illustration see Fig.\ \ref{fig2}. 

This BP reaches all the vertices of CDT triangulation and to
each CDT triangulation there is a distinct BP. Similarly
to each BP one can construct a  distinct CDT triangulation. 
The partition functions of 
the ensembles of BP and CDT are thus identical  except for a trivial 
redefinition of $\m$. 

From this identification it is clear that one can add a matter
system on the CDT triangulation which changes the critical
behavior of the triangulation, namely one can add hard dimers 
on the BP associated with the CDT triangulation. 
Adding such dimer systems to the CDT ensemble will  change the 
Hausdorff dimension of the CDT ensemble at the multicrititical
dimer point, such that a $m$-multicritical dimer system, $m=3,4,\ldots$
leads to a CDT random surface system with Hausdorff dimension
\rf{1.5}. That the Hausdorff dimension of the CDT ensemble is larger than
or equal to that of the BP ensemble is clear, since there
are more link-paths connecting two points in a CDT triangulation
than in the corresponding   
the BP-ensemble. If we assume there is a unique fractal 
structure of the CDT ensemble at all scales at the critical point,
it also has to agree with the fractal dimension assigned to the 
BP. This is because  the shortest link distance from the root of a BP to 
any vertex is essentially 
the same  as in the corresponding CDT triangulation.

These dimer 
models are of course somewhat artificial viewed from the perspective
of the two-dimensional CDT random triangulation. The rule 
for putting down the dimers (apart from being hard) 
is that you are not allowed to 
put them down on spatial links and for the links pointing 
forward in time from a vertex you are not allowed to put the 
dimer on the link furthest to the left. Nevertheless one would 
expect the critical behavior of a fullfledge hard dimer model 
on the random CDT surfaces to have the same critical behavior.
This is what we will show below using the matrix model representation
of a generalized CDT model introduced in \cite{newmatrix}.
 
\section{The multicritical CDT matrix models}

\subsection{Plain CDT}

\begin{figure}[t]
\centerline{\scalebox{0.6}{\includegraphics{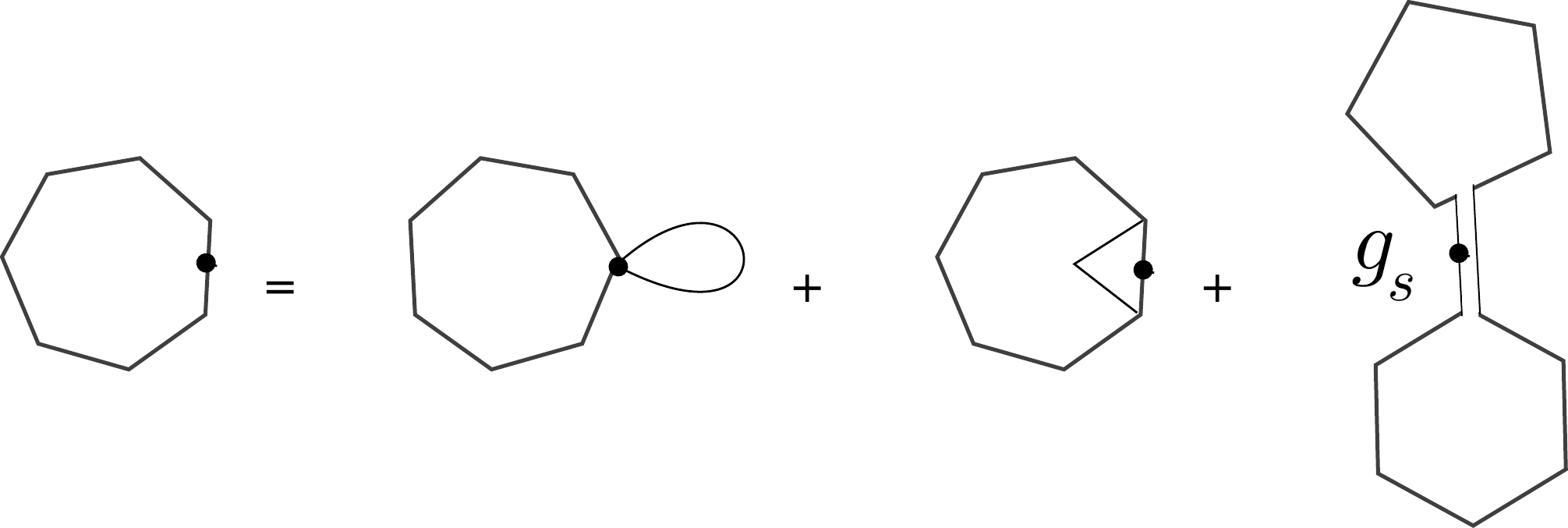}}}
\caption{The selfconsistent loop equation to be solved.
The coupling constant $g_s$ associated with the double link 
monitors the tendency for the creation of baby universes.}
\label{fig3}
\end{figure}

The ``plain'' CDT model can be obtained as limit of an
ordinary matrix model in the following way: 
assume the following matrix potential 
\beq\label{2.1}
V(\phi) =  \,\oh \phi^2 -
\lam \phi -  \frac{\lambda}{3} \phi^3 ,
\eeq
where the linear part of the potential is  just for convenience 
of the scaling limit (see \cite{newmatrix} for a discussion), and 
where $\phi$ is a $N\times N$ matrix. 
We then consider the following partition function:
\beq\label{2.2}
Z(\lam,g_s) = \int d\phi \; \e^{-\frac{N}{g_{s}} \tr V(\phi)}.
\eeq
Expanding the potential in powers of $\lam$ and performing 
the Gaussian integration results in an ensemble of 
random surfaces obtained by gluing together triangles 
(corresponding to the cubic term) and ``tadpoles'' corresponding
to the linear term. The logarithm of $Z$ represents the sum over
the connected surfaces, and taking the large $N$ limit will 
single out the surfaces of spherical topology. We will be 
interested in calculating the disk-amplitude, i.e.\ 
planar triangulations with a boundary. The corresponding 
object is 
\beq\label{2.3}
W(x) = \frac{1}{N} \la \tr \frac{1}{x-\phi} \ra,
\eeq
where the expectation value is wrt the partition function \rf{2.2}.
In the large $N$ limit the disk amplitude $W(x)$ satisfies 
the so-called loop equation
\beq\label{2.4}
g_s W(x)^2 = V'(x) W(x) -Q(x),~~~Q(x) = c_1 x +c_0,
\eeq 
where $V'$ denotes the derivative with respect to $x$.
The parameter $x$ is related to a boundary cosmological 
constant $\m_b$ by $x= e^{\m_b}$. The graphic representation
of the equation is shown in Fig. \ref{fig3}.

The solution is 
\beq\label{2.5}
W(x) = \frac{V'(x) -\sqrt{V'(x)^2 -4g_s Q(x)}}{2g_s},
\eeq
and the constants $c_0$ and $c_1$ are determined by the 
requirement that the $W(x)$ has a single cut on the real 
axis and that $W(x) = 1/x+ O(1/x^2)$ for $x \to \infty$.
Thus $W(x)$ has the following form
\beq\label{2.6}
W(x) =  \frac{V'(x) +\lam (x-c)\sqrt{(x-b)(x-a)}}{2g_s}.
\eeq
In this formula $a,b$ and $c$ are functions of the 
coupling constants $\lam$ and $g_s$.
The continuum limit where $W(x)$ can be associated with 
2d Euclidean quantum gravity is, for a fixed $g_s$,
the limit where  $c(\lam) \to b(\lam)$. In the 
neighborhood of this point, $\lam_c$, one has the expansion
\beq\label{2.7}
\lam = \lam_c(1 -\ep^2) \Lam,~~~b(\lam) = b(\lam_c)- \ep \sqrt{\Lam},  
~~~c(\lam) = c(\lam_c)  +\oh \ep \sqrt{\Lam},
\eeq
where 
\beq\label{2.8}
x=x_c+\ep X,~~~x_c= c(\lam_c) = b(\lam_c).
\eeq
The interpretation is that $\Lam$ is the continuum  
cosmological constant, $X$ the continuum boundary 
cosmological constant. In this limit the term $V'(x)$ does 
not scale and one obtains
\beq\label{2.9}
W(x) = {\it NS}(x) + {\rm const.}\; \ep^{3/2} W_{cont}(X),
\eeq
\beq\label{2.9a}
W_{cont}(X) = (X-\sqrt{\Lam}/2) \sqrt{X + \sqrt{\Lam}},
\eeq
where $W_{cont}(X)$ agrees with the disk amplitude calculated 
using quantum Liouville theory and ${\it NS}(x)$  is a non-scaling part,
analytic in $x$.

It is however possible to take another scaling limit 
related to CDT when $g_s \to 0$, more specifically we 
have to scale $g_s$ as follows
\beq\label{2.10}
g_s = G_s \ep^3.
\eeq
In this limit, which was denoted the ``classical'' limit, 
the behavior of the critical couplings to lowest order in $\ep$ is
\beq\label{2.11}
\lam_c(g_s) = \oh -\frac{3}{4} G_s^{2/3} \,\ep^2,~~~~
x_c(g_s) = 1 + G_s^{1/3}\,\ep,
\eeq
and the continuum cosmological constant and boundary cosmological constant
are defined as 
\beq\label{2.12}
\lam = \lam_c(g_s) -\ep^2 \,\Lam,~~~x=x_c(g_s) +\ep\, X. 
\eeq
Contrary to the situation for a fixed $g_s$, $W(x)$ itself 
will now scale and one obtains \cite{newmatrix}:
\beq\label{2.13}
W(x) = \ep^{-1}W_{cont}(X),
\eeq
\beq\label{2.13a}
W_{cont}(X) = \frac{\Lam_{{\rm cdt}} -\oh X_{{\rm cdt}}^2 + (X_{{\rm cdt}}-H)
\sqrt{(X_{{\rm cdt}} +H)^2 -\frac{4G_s}{H}}}{2G_s},
\eeq
where 
\beq\label{2.14}
\Lam_{{\rm cdt}} = \Lam +\frac{3}{2} G_s^{2/3},~~~~
X_{{\rm cdt}} = X + G_s^{1/3},
~~~2\Lam_{{\rm cdt}} H -H^3=2G_s.
\eeq
We note for future reference that the critical points 
\beq\label{2.14a}
(\lam_*,x_*) := (\lam_c(g_s=0),x_c(g_s=0))=(1/2,1)
\eeq
are characterized by 
\beq\label{2.14b}
V'(\lam_*,x_*)=V''(\lam_*,x_*)=0.
\eeq

In the limit where $G_s \to 0$ we obtain precisely the CDT disk function:
\beq\label{2.15}
W_{cont}(X) \to \frac{1}{X +\sqrt{2\Lam}} = \frac{1}{X} - 
\frac{\sqrt{2\Lam}}{X^2} + \cdots.
\eeq
$X \to \infty$ corresponds to contracting the boundary of the 
disk to a (marked) point, and the leading singularity of the 
corresponding closed (spherical) surface is thus $\sqrt{\Lam}$,
or re-introducing the bare coupling constants, $(\lam -\lam_c(g_s=0))^{1/2}$.
This critical behavior is precisely the same as the critical
behavior of the generic BP behavior, in agreement with the 
identification of the ensemble of CDT random surfaces with an 
ensemble of generic BPs mentioned above.

\subsection{Multicritical ``classical'' matrix models}

We now generalize the construction of a classical limit 
for ``plain'' CDT to the multicritical case. We consider the 
matrix potential 
\beq\label{2.16}
V(\phi) = \oh \phi^2 -
\lam \phi - \frac{\lam}{3} \phi^3- \frac{\lam^3\xi}{2} \phi^4.
\eeq
Viewed as a generating potential for random surfaces this 
matrix potential will glue together triangles and squares. 
Viewing each square as two triangles, we can think of the 
squares as part of the triangulation, but with a dimer 
placed on the diagonal, with a fugacity $\tilde{\xi}= \lam \xi$. In this 
way the model describes dimers put on  random triangulations
in a special way, such that there is at most one dimer per triangle.
On the graph dual to the triangulation the dimers are precisely 
hard dimers, and we will call them hard dimers also on the 
triangulation, even if the rule of putting down the dimers is 
slightly different from the standard hard dimer rule. Similarly 
we will denote $\xi$ the fugacity of the dimers, although it is 
strictly speaking $\tilde{\xi}$ which serves as the fugacity.

We are interested in a multicritical behavior of the matrix model
\rf{2.16} and it occurs for negative fugacity $\xi$. 
The disk amplitude is still given by \rf{2.5} where the 
polynomial $Q(x)$ now is of second order with coefficients uniquely 
fixed by the requirement that we have a single cut on the real 
line and that $W(x)$ falls off like $1/x$ for large $x$. The multicritical 
point $\lam_c(g_s),\xi_c(g_s)$ is characterized by the following behavior
of $W(x)$, generalizing the critical behavior above:
\beq\label{2.16a}
W(x) = \frac{V'(x) -2\lam^2_c(g_s)\xi_c(g_s) (x-b_c(g_s))^2 
\sqrt{(x-c_c(g_s))(x-a_c(g_s)})}{2g_s}.
\eeq
For a fixed $g_s$ the potential $V'(x)$ does not scale and 
plays no role in the continuum limit.

We are interested in a ``classical'' limit of \rf{2.16a} 
where the potential scales,
and that is obtained by the assignment
\beq\label{2.17}
g_s = G_s \ep^4.
\eeq 
Note that the scaling in \rf{2.17} differs from that in \rf{2.10}. This 
reflects that the multicritical point in the limit $g_s=0$ is characterized 
by the generalization of \rf{2.14b}:
\beq\label{2.18}
V'(\lam_*,\xi_*,x_*)=V''(\lam_*,\xi_*,x_*)= V'''(\lam_*,\xi_*,x_*)=0.
\eeq
From  \rf{2.16} and \rf{2.16a} it follows that 
\beq\label{2.19}
x_*= \frac{1}{\lam_*}= -\frac{1}{2\xi_*}= \sqrt{3}.
\eeq                                   
Thus a canonical  scaling of the boundary cosmological constant 
around $x_*$ will lead to a $V'(\lam_*,\xi_*,x)$ of order $\ep^3$
and a behavior compatible with \rf{2.13} is obtained. 

With $g_s$ scaling like \rf{2.17} one can calculate 
$\lam_c(g_s),\xi_c(g_s),x_c(g_s)$:
\beq\label{2.20}
\lam_c(g_s) = \lam_*\Big(1 - \frac{\sqrt{5}}{9} G_s^{1/2} \ep^2-
\frac{16\sqrt{5}}{27} G_s^{3/4}\ep^3\Big) + O(\ep^4),
\eeq
\beq\label{2.21}
\xi_c(g_s)= \xi_*\Big(1 - \frac{\sqrt{5}}{9} G_s^{1/2} \ep^2+
\frac{16\sqrt{5}}{27} G_s^{3/4}\ep^3\Big) + O(\ep^4),
\eeq
\beq\label{2.22}
x_c(g_s) = b_c(g_s)= x_*\Big(1+\frac{2}{3\,5^{1/4}} G_s^{1/4}\ep\Big) 
+O(\ep^2).
\eeq

The perturbation away from $\lam_c(g_s),\xi_c(g_s)$ which leads
to a potential $V'(x,\lam,\xi)$ of order $\ep^3$, assuming 
the boundary cosmological constant is perturbed as $x=x_c(g_s)+\ep X$,
can be parametrized as 
\beq\label{2.23}
\lam = \lam_* +\tilde{\Lam} \ep^2-\Lam \, \ep^3,~~~
\xi=\xi_* -\oh \tilde{\Lam} \ep^2.
\eeq
As in the ordinary multicritical matrix model situation 
one finds a two-parameter set of solutions depending on $\Lam,\tilde{\Lam}$,
rather than the solution \rf{2.13a} which only depends on one paramter, 
$\Lam$. Let us choose a convenient ``background'', using the notation
from ordinary matrix models \cite{sei-stau}, which we call the 
CDT-background, namely $\tilde{\Lam}=0$. 
By a redefinition similar to \rf{2.14}:
\beq\label{2.23a} 
\Lam_{\rm cdt} = \Lam + \frac{32 \sqrt{3} \,5^{1/4}}{81} G_s^{3/4},~~~
X_{\rm cdt} = X + \frac{2}{\sqrt{3}\,5^{1/4}}G_{s}^{1/4},
\eeq
we obtain
\beq\label{2.24}
V'(x)= \ep^3 \Big( \Lam_{\rm cdt} + \frac{1}{9} X_{\rm cdt}^3\Big).
\eeq
One can now calculate $W(x)$ in the CDT limit $G_s \to 0$ where 
any creation of baby universes is suppressed:
\beq\label{2.25}
W(x)= \frac{1}{\ep} \frac{1}{  X_{\rm cdt} + \Lam_{\rm cdt}^{1/3}}.
\eeq
Repeating the exercise below eq.\ \rf{2.15}, taking $X_{\rm cdt}$ to 
infinity, we obtain that the critical behavior of the partition function
of spherical multicritical CDT surfaces with one marked point is
dominated by the term $(\lam-\lam_c(g_s=0))^{1/3}$, in agreement 
with the expectation we have from the multicritical BP model.

Finally, let us note that it follows from \rf{2.23} that the 
critical exponent $\sg$ of the hard dimers is 1/2, again in complete
analogy with the multicritical BP model calculation \rf{1.8}-\rf{1.10}.   
The result $\sg = 1/2$ is also the result obtained for ordinary 
multicritical matrix models. In this sense it is not really surprising 
that we obtain the result. One obtains the CDT multicritical point 
by following the ordinary multicritical line $\lam_c(g_s),\xi_c(g_s)$
from, say, $g_s=1$ to the CDT value $g_s=0$. For any fixed value of 
$g_s$ we have of course $\sg=1/2$, and although $\sg$ might have 
changed when taking the particular classical scaling limit $g_s = \ep^4 G_s$,
this is apparently not what happens.

\section{Discussion}   
   
Only a few riddles are left in 2d Euclidean quantum gravity
coupled to matter. One of them is the behavior of the Hausdorff 
dimension $d_h$ as a function of the central charge $c$ of the 
conformal theory coupled gravity. A formula was derived by Watabiki
some years ago \cite{watabiki}
\beq\label{3.1}
d_h = 2 \frac{\sqrt{49-c}+\sqrt{25-c}}{\sqrt{25-c}+\sqrt{1-c}}
\eeq
Most likely this formula is correct for $c\leq 0$. For $c=0$ agrees
with what is known to be the correct answer \cite{dh4}. For $c=-2$
there are very reliable computer simulations  
which show agreement with the formula
\cite{c-2}. Finally for $c \to -\infty$ it gives  2, something one would 
indeed expect from semiclassical Liouville theory. However, 
for $0<c\leq 1$ the numerical agreement is less conclusive
\cite{DTisingsimu}, and 
the possibility that $d_h=4$ in this range was pointed out, and 
the idea has recently been resurrected \cite{dupliant}. 
For $c>1$ the Liouville    
formulas become complex and expressions like \rf{3.1} are not 
valid, but it is believed that there is a universal phase where the world
sheet degenerates to BPs. 

Surprisingly we have a somewhat similar situation in CDT: from numerical 
simulations $d_h$ seems unchanged (and equal 2) when matter with
central charge $0\leq c \leq 1$ is coupled to the CDT ensemble 
\cite{2dcdtsimu} and recently it was shown that there might
be a kind of universal phase for $c>1$ \cite{clarge}.
However, to the extent we can really view the hard dimer models 
as corresponding to conformal field theories, it seems that for $c<0$ 
the matter systems can change fractal structure of the CDT ensemble.
Qualitatively the changes are like in the full Euclidean models, 
$d_h$ decreases with decreasing $c$. In the $c=0$ case the CDT model
can be understood as an effective Euclidean model, where baby 
universes have been integrated out. Whether such an interpretation
is possible also when  matter is coupled to CDT is presently unknown, 
but since the multicritical matrix models capture the critical 
behavior of both CDT and ordinary 2d  Euclidean quantum gravity coupled
to certain matter systems, depending on how we scale $g_s$, we 
have a chance to answer this question in the context of matrix models.

\subsection*{Note}

While completing this article we were informed by Stefan Zohren that
he and Max Atkin have obtained results which are identical to 
a number of our results. We thank Stefan for informing us of these results
prior to publication.
 
\vspace{.5cm}      
\noindent {\bf Acknowledgements.} 
JA thanks Vincent Bonzom and Stefan Zohren for discussions.
JA would like to thank the Institute of 
Theoretical Physics and
the Department of Physics and Astronomy at Utrecht University for hospitality 
and financial support. JA also thanks the Perimeter Istitute for hospitality
and financial support. JA, LG and AG thanks the Danish Research Council 
for financial support via the grant ``Quantum gravity and the role
of Black holes''. 
AG acknowledges a partial support by the Polish Ministry of 
Science grant N N202 229137 (2009-2012).
YS would like to thank Hiroyuki Fuji for useful conversations.
YS also thanks the Niels Bohr Institute for  warm hospitality.
YS is supported by the Grant-in-Aid for Nagoya University Global COE 
Program, "Quest for Fundamental Principles in the Universe: from Particles to
the Solar System and the Cosmos".


\begin{thebibliography}{99}

\bibitem{kazakov}
  V.~A.~Kazakov,
  Phys.\ Lett.\  {A\ 119} (1986)  140-144.
  
\bibitem{staudacher}
  M.~Staudacher,
  Nucl.\ Phys.\ B {\bf 336} (1990) 349.



\bibitem{adj-bp}
  J.~Ambjorn, B.~Durhuus and T.~Jonsson,
  Mod.\ Phys.\ Lett.\ A {\bf 9} (1994) 1221
  [hep-th/9401137].

\bibitem{BB}
  P.~Bialas and Z.~Burda,
  Phys.\ Lett.\ B {\bf 384} (1996) 75,
  [hep-lat/9605020].



\bibitem{djf}
  B.~Durhuus, J.~Frohlich and T.~Jonsson,
  Nucl.\ Phys.\ B {\bf 240} (1984) 453;
  Phys.\ Lett.\ B  {\bf 137} (1984) 93.



\bibitem{DT}
  F.~David,
  Nucl.\ Phys.\  {\bf B257 } (1985)  45.\\
A.~Billoire and F.~David,
Phys.\ Lett.\  B\ {\bf 168} (1986) 279-283.\\
J.~Ambjorn, B.~Durhuus and J.~Fr\"ohlich,
Nucl.\ Phys.\  B\ {\bf 257} (1985) 433-449;\\
J.~Ambjorn, B.~Durhuus, J.~Fr\"ohlich and P.~Orland,
Nucl.\ Phys.\  B\ {\bf 270} (1986) 457-482.\\
  V.~A.~Kazakov, A.~A.~Migdal, I.~K.~Kostov,
  Phys.\ Lett.\  {\bf B157 } (1985)  295-300.\\
D.V.~Boulatov, V.A.~Kazakov, I.K.~Kostov and A.A.~Migdal,
Nucl.\ Phys.\  B\ {\bf 275} (1986) 641-686.

\bibitem{ad}
  J.~Ambjorn and B.~Durhuus,
  Phys.\ Lett.\ B {\bf 188} (1987) 253.



\bibitem{higherD}
J.~Ambjorn and J.~Jurkiewicz,
Phys.\ Lett.\  B\ {\bf 278} (1992) 42-50.\\
M.E.~Agishtein and A.A.~Migdal,
Mod.\ Phys.\ Lett.\  A\ {\bf 7} (1992) 1039-1062.\\
 J.~Ambjorn, S.~Jain, J.~Jurkiewicz, C.F.~Kristjansen,
  Phys.\ Lett.\  {B\ 305} (1993)  208-213
  [hep-th/9303041].
  

\bibitem{CDT-highD}
J.~Ambjorn, A.~G\"orlich, J.~Jurkiewicz and R.~Loll,
Phys.\ Rev.\ Lett.\ {\bf 100} (2008) 091304 [arXiv:0712.2485, hep-th].\\
Phys.\ Rev.\  D {\bf 78} (2008) 063544 [arXiv:0807.4481, hep-th].\\
J.~Ambjorn, J.~Jurkiewicz and R.~Loll,
Phys.\ Lett.\ B {\bf 607} (2005) 205-213
[hep-th/0411152].\\
Phys.\ Rev.\ Lett.\ {\bf 93} (2004) 131301 [hep-th/0404156].\\
Phys.\ Rev.\ D\ {\bf 72} (2005) 064014  [hep-th/0505154].



\bibitem{al}
  J.~Ambjorn, R.~Loll,
  Nucl.\ Phys.\  {B\ 536} (1998)  407-434
  [hep-th/9805108].
 


\bibitem{ackl}
  J.~Ambjorn, J.~Correia, C.~Kristjansen, R.~Loll,
  Phys.\ Lett.\  {B\ 475} (2000)  24-32
  [hep-th/9912267].

\bibitem{djw}
  B.~Durhuus, T.~Jonsson and J.~F.~Wheater,
  arXiv:0908.3643 [math-ph].



\bibitem{newmatrix}
J.~Ambjorn, R.~Loll, W.~Westra and S.~Zohren,
JHEP {\bf 0712} (2007) 017 [arXiv:0709.2784, gr-qc];\\
Phys.\ Lett.\  B\ {\bf 665} (2008) 252-256 [arXiv:0804.0252, hep-th];\\
Phys.\ Lett.\  B {\bf 670} (2008) 224 [arXiv:0810.2408, hep-th].\\
JHEP {\bf 0805} (2008) 032 [arXiv:0802.0719, hep-th].

\bibitem{book}
  J.~Ambjorn, B.~Durhuus and T.~Jonsson,
{\it Quantum geometry. A statistical field theory approach,}
  Cambridge, UK: Univ. Pr., 1997. 
(Cambridge Monographs in Mathematical Physics). 363 p


\bibitem{bonzom}
  V.~Bonzom,
  arXiv:1201.1931 [hep-th].




\bibitem{durhuus}
  B.~Durhuus and G.~M.~Napolitano,
  arXiv:1107.2964 [cond-mat.stat-mech].



\bibitem{adf3}
  J.~Ambjorn, B.~Durhuus and J.~Frohlich,
  Nucl.\ Phys.\ B {\bf 275} (1986) 161.





\bibitem{sei-stau}
  G.~W.~Moore, N.~Seiberg and M.~Staudacher,
  Nucl.\ Phys.\ B {\bf 362} (1991) 665.


\bibitem{watabiki}
  Y.~Watabiki,
  Prog.\ Theor.\ Phys.\ Suppl.\  {\bf 114} (1993) 1.


\bibitem{dh4}
H.~Kawai, N.~Kawamoto, T.~Mogami, Y.~Watabiki,
Phys.\ Lett.\ B\ 306 (1993) 19-26
[hep-th/9302133].\\
J.~Ambjorn, Y.~Watabiki,
Nucl.\ Phys.\ B\ 445 (1995) 129-144 [hep-th/9501049].
H.~Aoki, H.~Kawai, J.~Nishimura, A.~Tsuchiya,
Nucl.\ Phys.\ B\ 474 (1996) 512-528 [hep-th/9511117].






\bibitem{c-2}
  J.~Ambjorn, K.N.~Anagnostopoulos, T.~Ichihara, L.~Jensen, N.~Kawamoto, Y.~Watabiki, K.~Yotsuji,
  Phys.\ Lett.\  {B\ 397} (1997)  177-184
  [hep-lat/9611032];
  Nucl.\ Phys.\  {B\ 511} (1998)  673-710
  [hep-lat/9706009].


\bibitem{DTisingsimu}
  J.~Ambjorn and K.~N.~Anagnostopoulos,
  Nucl.\ Phys.\ B {\bf 497} (1997) 445
  [hep-lat/9701006].\\
  J.~Ambjorn, K.~N.~Anagnostopoulos, U.~Magnea and G.~Thorleifsson,
  Phys.\ Lett.\ B {\bf 388} (1996) 713
  [hep-lat/9606012].


\bibitem{dupliant}
  B.~Duplantier,
  arXiv:1108.3327 [math-ph].




\bibitem{2dcdtsimu}
J.~Ambjorn, K.N.~Anagnostopoulos, R.~Loll and I.~Pushkina:
Nucl.\ Phys.\  B {\bf 807} (2009) 251 [arXiv:0806.3506, hep-lat].\\
  J.~Ambjorn, K.~N.~Anagnostopoulos and R.~Loll,
  Phys.\ Rev.\ D {\bf 60} (1999) 104035
  [hep-th/9904012].\\
  J.~Ambjorn, K.~N.~Anagnostopoulos and R.~Loll,
  Phys.\ Rev.\ D {\bf 61} (2000) 044010
  [hep-lat/9909129].

\bibitem{clarge}
  J.~Ambjorn, A.~T.~Goerlich, J.~Jurkiewicz and H.~G.~Zhang,
  arXiv:1201.1590 [gr-qc].




\end{thebibliography}
\end{document}